\documentclass[journal]{IEEEtran}
\IEEEoverridecommandlockouts
\usepackage{amsmath,amssymb,amsfonts}
\usepackage{algorithmic}
\usepackage{graphicx}
\usepackage{textcomp}
\usepackage{xcolor}
\usepackage{silence}
\usepackage{subcaption}
\usepackage{adjustbox}
\usepackage{url}
\usepackage{tikz}
\usepackage{pgfplots}
\pgfplotsset{compat=1.7}
\usepackage{numprint}

\usepackage{mathdots}
\usepackage{yhmath}
\usepackage{cancel}
\usepackage{color}
\usepackage{siunitx}
\usepackage{array}
\usepackage{multirow}
\usepackage{gensymb}
\usepackage{tabularx}
\usepackage{booktabs}
\usetikzlibrary{fadings}
\usetikzlibrary{patterns}
\usetikzlibrary{shadows.blur}
\usetikzlibrary{shapes}

\def\BibTeX{{\rm B\kern-.05em{\sc i\kern-.025em b}\kern-.08em
    T\kern-.1667em\lower.7ex\hbox{E}\kern-.125emX}}

\begin{document}

\title{\LARGE A Scalable Architecture for Monitoring IoT Devices Using Ethereum and Fog Computing
\thanks{978-1-7281-9611-4/20/\$31.00~\copyright2020~IEEE}
}

\author{\IEEEauthorblockN{Shirin Tahmasebi, Jafar Habibi and Abolhassan Shamsaie}\\
\vspace*{0.5cm}
\IEEEauthorblockA{\small{\textit{Computer Engineering Department}, \textit{Sharif University of Technology}, Tehran, Iran, 
shtahmasebi@ce.sharif.edu\\
\textit{Computer Engineering Department}, \textit{Sharif University of Technology}, Tehran, Iran, 
jhabibi@sharif.edu\\
\textit{Computer Engineering Department}, \textit{Sharif University of Technology}, Tehran, Iran, 
shamsaie@ce.sharif.edu}}}


\IEEEabstract{
With recent considerable developments in the Internet of Things (IoT), billions of resource-constrained devices are interconnected through the Internet. Monitoring this huge number of IoT devices which are heterogeneous in terms of underlying communication protocols and data format is challenging. The majority of existing IoT device monitoring solutions heavily rely on centralized architectures. Since using centralized architectures comes at the expense of trusting an authority, it has several inherent drawbacks, including vulnerability to security attacks, lack of data privacy, and unauthorized data manipulation. Hence, a new decentralized approach is crucial to remedy these drawbacks. One of the most promising technologies which is widely used to provide decentralization is blockchain. Additionally, to ease the burden of communication overhead and computational power on resource-constrained IoT devices, fog computing can be exploited to decrease communication latency and provide better network scalability.

In this paper, we propose a scalable blockchain-based architecture for monitoring IoT devices using fog computing. To demonstrate the feasibility and usability of the proposed solution, we have implemented a proof-of-concept prototype, leveraging Ethereum smart contracts. Finally, a comprehensive evaluation is conducted. The evaluation results indicate that the proposed solution is significantly scalable and compatible with resource-constrained IoT devices.
}

\IEEEkeyword{
Internet of things, device monitoring, fog, blockchain, smart contract, ethereum, ipfs
}

\maketitle

\section{Introduction}

The Internet of Things (IoT) is a network of connected physical devices. The ability to connect billions of devices and the inherent benefits of IoT has led to remarkable growth in IoT-based services. The primary role of IoT devices is to detect and measure changes in their surroundings and provide context-awareness~\cite{introduction_centralizationDisadvantages_1}. By leveraging the generated data, a variety of services can be delivered to users. The quality of these delivered services is highly dependent on the quality and correctness of the operations of devices~\cite{introduction_centralizationDisadvantages_1}. Hence, it is of pivotal importance to monitor IoT devices and be assured of their normal operations~\cite{introduction_monitoring}.


Several solutions have been proposed for monitoring IoT devices. However, the majority of them have relied on traditional centralized architectures and client-server models. Using centralized architecture has some critical disadvantages, including:

\begin{itemize}
    \item These solutions are based on trusting a centralized entity that has full authority over the system~\cite{introduction_centralizationDisadvantages_1, introduction_centralizationDisadvantages_2}. Trusting a centralized third-party with full authority makes the system prone to data manipulation and unauthorized data sharing~\cite{introduction_centralizationDisadvantages_1, introduction_centralizationDisadvantages_3}.
    \item A centralized infrastructure can introduce a single point of failure, which can potentially degrade the availability level and reduce the Quality of Service (QoS)~\cite{introduction_centralizationDisadvantages_1, introduction_centralizationDisadvantages_2}.
    \item Centralized architectures are vulnerable to a variety of security threats, such as DDoS/DoS attack~\cite{introduction_centralizationDisadvantages_1, introduction_centralizationDisadvantages_3}.
\end{itemize}


Overcoming these problems requires proposing more suitable solutions. These solutions need to be fully distributed and be able to provide a completely trustless environment; wherein no single entity has authority over the system. In this regard, blockchain - a distributed, decentralized, immutable, and public digital ledger that records transactions across a peer-to-peer network - seems to be a promising technology~\cite{bitcoin_whitePaper,ethereum_whitePaper}. Moreover, blockchain provides an entirely trustless environment and can be leveraged to overcome the problems of centralized architectures. In addition, blockchain enables a self-regulated and self-managed environment by leveraging smart contracts (self-executing code run by mining nodes).
Recently, blockchain technology has attracted considerable attention from many researchers, and many studies have investigated the benefits of using blockchain in different IoT domains. For instance, in \cite{introduction_blockchainIot_1} and \cite{introduction_blockchainIot_2}, the authors studied the challenges of using blockchain in the IoT context. In \cite{introduction_blockchainIot_3}, \cite{introduction_blockchainIot_4}, and \cite{introduction_blockchainIot_5}, the feasibility of using blockchain in different scenarios of smart cities is investigated. Similarly, several research papers addressed applying the combination of IoT and blockchain in supply chain management, including \cite{introduction_blockchainIot_6}, \cite{introduction_blockchainIot_7}, and \cite{introduction_blockchainIot_8}.

Successfully adopting blockchain to the IoT domain necessitates considering:
\begin{itemize}
    \item Many IoT devices face serious limitations in terms of processing power, storage, and battery resources and are not capable of performing resource-intensive operations.
    
    \item A typical IoT network usually comprises of a massive number of IoT devices that sporadically transmit and broadcast data. In addition, IoT devices are heterogeneous in terms of underlying communication protocols, data formats, and technologies. Hence, scalability is an undeniable need for IoT solutions.

\end{itemize}
Therefore, it is crucial to take these considerations into account while designing blockchain-based IoT solutions. However, there is a main concern, referred to as blockchain scalability trilemma, that potentially can impede satisfying these considerations. This concept was coined by Vitalik Buterin - founder of Ethereum - to describe the blockchain limitation in addressing scalability, decentralization, and security, simultaneously and without comprising any of them~\cite{introduction_blockchainTrilemma}. Therefore, it is a challenge to handle the growing amount of transactions in a blockchain.

Moreover, another promising technology that can help to better address scalability and latency requirements is fog computing. Fog computing transfers computational and communication capabilities close to IoT devices to decrease latency, distribute storage resources, improve mobility, provide context-awareness, and finally to boost network scalability. Hence, fog computing can be used to ease the burden on resource-constrained IoT devices. \cite{introduction_fogComputing}

In this paper, our objective is to design a distributed and scalable architecture for monitoring IoT devices using blockchain technology. Furthermore, to increase the usability of our solution in a variety of IoT scenarios, it is vital to make sure that our solution is specifically tailored to the resource-constrained nature of IoT devices. Additionally, our solution is designed based on Ethereum blockchain; a public and permissionless blockchain which provides a decentralized and Turing-complete virtual machine, named as Ethereum Virtual Machine (EVM), to execute smart contract scripts. Ethereum has been successfully applied in many cases. 

\noindent\textbf{Contributions.} Our major contributions are listed as follows:
\begin{itemize}
    \item We present a blockchain-based architecture for monitoring IoT devices, using Ethereum and fog computing.
    
    \item We take advantage of smart contracts to provide a mechanism to define dynamic monitoring policies.
    
    \item Our novel approach is specifically designed to be highly scalable and interoperable with resource-constrained IoT devices.
    
    \item We provide further details of possible scenarios in the proposed architecture.
    
    \item We provide a proof-of-concept implementation and develop a decentralized application (DApp). Moreover, we deploy the DApp on a test network.
    
    \item We evaluate our solution in terms of performance and scalability.
\end{itemize}

\noindent\textbf{Organization.} In Section~\ref{RelatedWork} a comprehensive review of related work is presented. In Section~\ref{ProposedSolution}, we describe our blockchain-based architecture and discuss why some architectural decisions are made to satisfy the essential IoT network requirements. In Section~\ref{evaluation}, we present a proof-of-concept implementation followed by a discussion on evaluation results. Moreover, we investigate the limitations and benefits of our proposed solution. Finally, Section~\ref{Conclusion} concludes the paper.

\section{Related Work}
\label{RelatedWork}

So far, several researchers have relied on centralized models to design IoT device monitoring solutions. An overview of these centralized solutions is provided in Section~\ref{RelatedWork_centralized}. Moreover, regarding the disadvantages of centralized models, blockchain technology, as a candidate solution to bring decentralization to the systems, has attracted extensive attention in recent years. As far as using blockchain technology in the IoT context is concerned, we investigate some use cases of combining these two technologies in Section~\ref{RelatedWork_decentralized}.

\subsection{Centralized Protocol-based Models}
\label{RelatedWork_centralized}

In the literature, there already exist several protocol-based models for monitoring IoT devices. In this section, we review some studies that have employed a variety of device communication protocols to present a centralized model for monitoring IoT devices. It is worth pointing out that although most of these studies are mainly based on centralized models, the communication protocols that are leveraged in these studies are also applicable in decentralized models.

Simple Network Management Protocol (SNMP) is a network management and monitoring protocol that is based on the manager/agent model. The exchange information between agents and managers is hierarchically organized.  This hierarchy is described in the Management Information Base (MIB) files~\cite{relatedWork_managementProtocols_survey}. Although SNMP was designed without taking into account resource-constrained devices, several papers highlighted the benefits of using the SNMP protocol for IoT device management and monitoring. In~\cite{relatedWork_snmp_1}, the authors investigated how to use an SNMP-based agent to monitor IoT devices. However, they did not conduct any evaluation. In~\cite{relatedWork_snmp_2}, the authors not only leveraged the SNMP protocol to manage and monitor IoT devices but also elaborated on the device control interface design to ease the communication with the SNMP-based agent. Moreover, since full compatibility of SNMP protocol with resource-constrained devices is indecisive, in~\cite{relatedWork_snmp_3}, the authors focused on designing a novel manager-agent model and conducted several experiments to reveal that their solution is compatible with limited devices.

Message Queuing Telemetry Transport (MQTT) is a protocol specifically designed to ease communication between resource-constrained devices. It uses a publisher/subscriber model and heavily relies on the existence of a mediator, referred to as broker, to handle the procedure of subscription and publication~\cite{relatedWork_managementProtocols_survey}. Using MQTT for IoT device monitoring has been widely investigated in several studies. In~\cite{relatedWork_mqtt_1}, a three-layer architecture is designed to monitor IoT devices using MQTT protocol, and a web-based interface is provided to visualize collected data. In~\cite{relatedWork_mqtt_2}, the authors proposed a system to collect instantaneous data from agricultural smart devices using MQTT. Then, the gathered data is fed to several machine learning algorithms to conduct some predictive analysis.

Another efficient protocol for monitoring resource-constrained devices is the Constrained Application Protocol (CoAP). CoAP is an application-level protocol that follows a client/server model~\cite{relatedWork_managementProtocols_survey}. In~\cite{relatedWork_coap_1}, CoAP is used to design a scalable architecture for resource discovery in IoT networks.  In~\cite{relatedWork_coap_2}, a CoAP-based system is proposed to collect medical data from devices. Moreover, to secure the communication, the Elliptic Curve Cryptographic algorithm is used. Finally, COOJA simulator is used to evaluate the system. In~\cite{relatedWork_coap_3}, the authors present an IoT device management system based on CoAP. Also, they designed a novel model to map resources to several predefined classes logically. 

The CPE WAN Management Protocol (CWMP), also known as TR-069, is an XML-based protocol for remote network device management. This protocol uses a client/server model in which the managed devices (referred to as CPE) and the server (referred to as Auto Configuration Server or ACS) can have bidirectional communication over HTTP~\cite{relatedWork_managementProtocols_survey}. Since CWMP has been successfully applied in a variety of commercial cases, several papers investigated whether it is feasible to use it in the IoT context.  In~\cite{relatedWork_cwmp_1}, CWMP is applied for the management of an IoT network. However, since it is assumed that the devices always have a proper Internet connection and are capable of exchanging data in XML format, it seems that their proposed approach is not applicable in a wide range of resource-constrained devices. In~\cite{relatedWork_cwmp_2}, the authors believe that CWMP, as it is, is merely feasible in resource-intensive devices. Hence, they presented a novel data model to enhance the compatibility of this protocol with IoT devices.

\textbf{Limitations.} To summarize this section, we have extensively investigated a number of papers that employed centralized models for IoT device monitoring and management. As mentioned previously, since centralized models rely on the existence of intermediate entities, the whole system, as well as users' data, is under the control of a single authority. This authority not only may misuse data but also can be vulnerable to security attacks. Moreover, centralized models can become a single point of failure, which negatively affects availability and QoS.

\subsection{Decentralized Blockchain-based Models}
\label{RelatedWork_decentralized}
In this section, we review some pieces of research that used blockchain technology to design a distributed IoT system. In~\cite{relatedWork_blockchain_1}, the authors presented a blockchain-based system for managing IoT devices. Furthermore, they used Ethereum smart contracts to store data coming from IoT devices. However, their proposed solution has critical weaknesses. First of all, it is assumed that devices are full nodes and are in direct communication with Ethereum blockchain. This assumption seems unreasonable concerning the resource-constrained nature of IoT devices. Secondly, it is considered that the generated data is directly stored on the Ethereum blockchain. Since storing data to the Ethereum blockchain emits a transaction and consumes gas, it can negatively affect the applicability of their proposed solution.
In~\cite{relatedWork_blockchain_3}, the authors presented a blockchain-based configuration system for network devices. In their system, all configuration files and the history of all configuration changes are persisted in Hyperledger Fabric. Moreover, they mentioned two limitations of their solution. First, it is assumed that network devices are blockchain peers. Second, all configuration information, including potentially large configuration files, is stored on the blockchain. Furthermore, they proposed an extended version of their solution in~\cite{relatedWork_blockchain_3_extended}. They aimed to optimize their architecture to be compatible with IoT networks. In their extended solution, IoT devices are not blockchain peers anymore, and also large configuration files are kept in an external database with only their signatures in the blockchain.
The authors in~\cite{relatedWork_blockchain_2}, provided a decentralized framework to store sensor data securely. Moreover, they implemented a proof-of-concept using Hyperledger Fabric.

\textbf{Limitations.}
Since the combination of blockchain technology in the IoT context is concerned, several works in this theme are studied. 
However, to the best of our knowledge, research works on decentralized monitoring solutions are limited.

\subsection{Research Gap}
To the best of our knowledge, the only previous work aimed to apply the blockchain technology to the context of monitoring IoT devices is~\cite{relatedWork_blockchain_3_extended}, which is an extended version of~\cite{relatedWork_blockchain_3}. However, the major difference between the work presented in~\cite{relatedWork_blockchain_3_extended} and ours is that our work focuses on making sure of the device normal and correct operation rather than just keeping track of configuration changes. Moreover, we provide dynamic monitoring of devices by taking advantage of defining monitoring policies.

\section{Proposed Solution}
\label{ProposedSolution}

Our novel proposed architecture describes a decentralized system for monitoring IoT devices using blockchain technology and fog computing. 
In the proposed solution, the following notes are considered: 
\begin{enumerate}
    \item
    Each user can register and add multiple devices. To add a device, the user inputs the device IP address, its model, its credentials, desired polling interval, target attributes to be collected, and so on.
    
    \item 
    Each IoT device can be equipped with several resources (such as processing units, storage, battery resources, and several IoT sensors).  Each resource has several attributes, which can potentially be a target for monitoring.
    
    \item  
    To monitor devices, we define some monitoring policies. A monitoring policy comprises several properties, including the attributes to be collected, their maximum and minimum thresholds, the allowed numbers of threshold violations, and the criticality level of the violations. 

    \item 
    According to monitoring policies, when the number of boundary violations for an attribute exceeds the allowed number of violations, a proper event with determined criticality level is emitted.

\end{enumerate}


To illustrate the proposed architecture, a layered architecture of the system is presented in section \ref{layeredArchitecture}. Then, a more detailed architecture is depicted in section \ref{conceptualModel}. Finally, section \ref{systemInteractions} describes the interactions and possible scenarios in the proposed architecture in more details.

\subsection{Layered Architecture}
\label{layeredArchitecture}

\begin{figure}[htbp]
    \centering
    \includegraphics[scale=0.40]{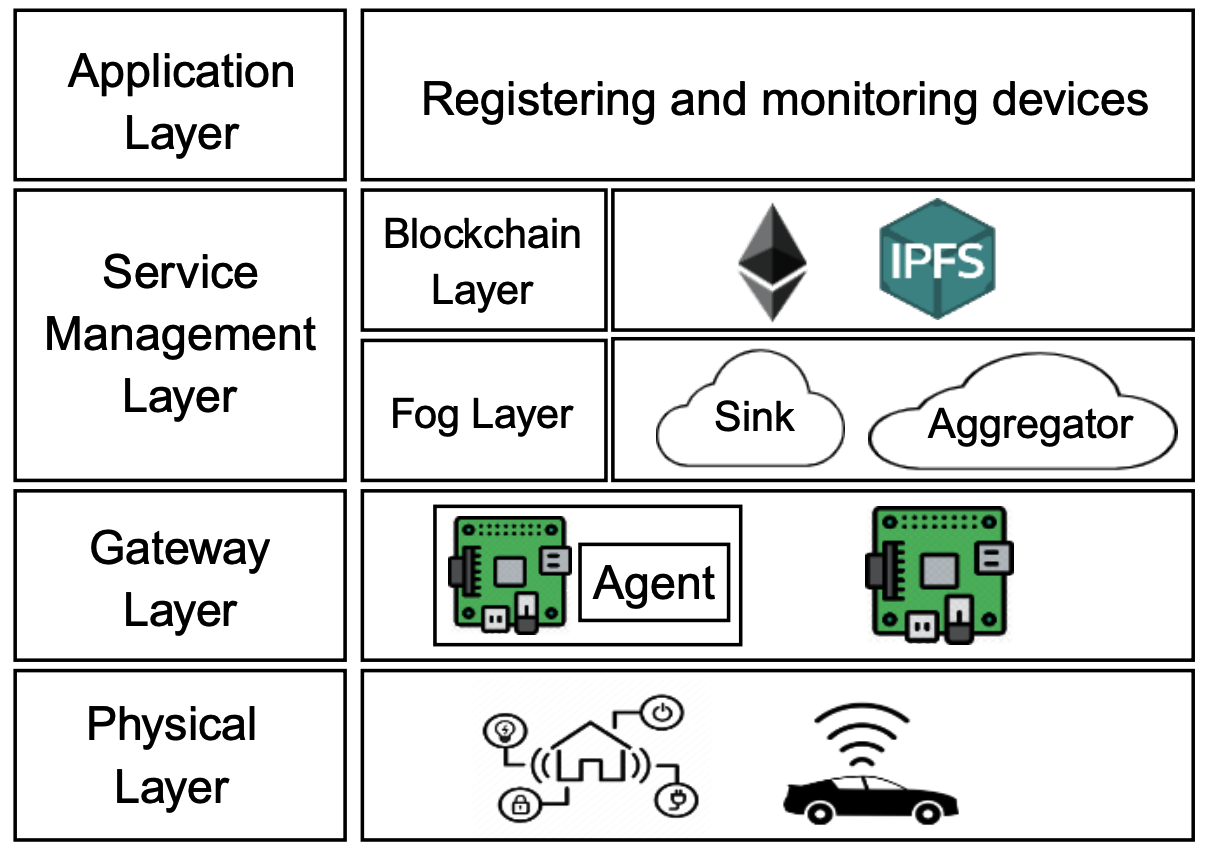}
    \caption{Proposed architecture layered model}
    \label{fig:layeredArchitecture}
\end{figure}

A four-layered architecture of our proposed method is presented in Figure~\ref{fig:layeredArchitecture} that is comprised of physical layer, gateway layer, service management layer, and application layer. Each layer is briefly described as follows:

\begin{enumerate}
    \item \textbf{Physical Layer}:
    This layer consists of basic hardware such as sensors and actuators. These are used to collect information from their context and then convert this information into digital data. The digitized data is then passed to the gateway layer.
    
    \item \textbf{Gateway Layer}:
    The devices which are used in this layer perform the task of collecting device data and preprocessing them. The preprocessed data is then passed to the service management layer. Each of these devices can be equipped with an external software agent that is used to have an integrated interface to ease communication.
    
    \item \textbf{Service Management Layer}:
    This layer consists of devices with more computing power, better data analytic features, and more storage capacity in comparison with those in the gateway layer. 
    This layer has the responsibility to:
    \begin{enumerate}
        \item 
        Store users' input about monitoring policies.
        
        \item
        Poll and collect data from gateway layer based on defined policies.
        
        \item
        Handle the huge amount of data coming from beneath layers to emit events. Events are generated according to monitoring policies.
        
    \end{enumerate}
    
    Furthermore, this layer is comprised of two sub-layers: blockchain layer and fog layer, each of which will be explained in more detail in the following sections.

    \item \textbf{Application Layer}:
    The application layer provides various services and applications to the users or the customers according to the processed data in the service management layer. Applications such as registering and monitoring the devices are provided in this layer.
    
\end{enumerate}

\subsection{System Architecture}
\label{conceptualModel}

\begin{figure*}[htb!]
    \centering
    \includegraphics[scale=0.7]{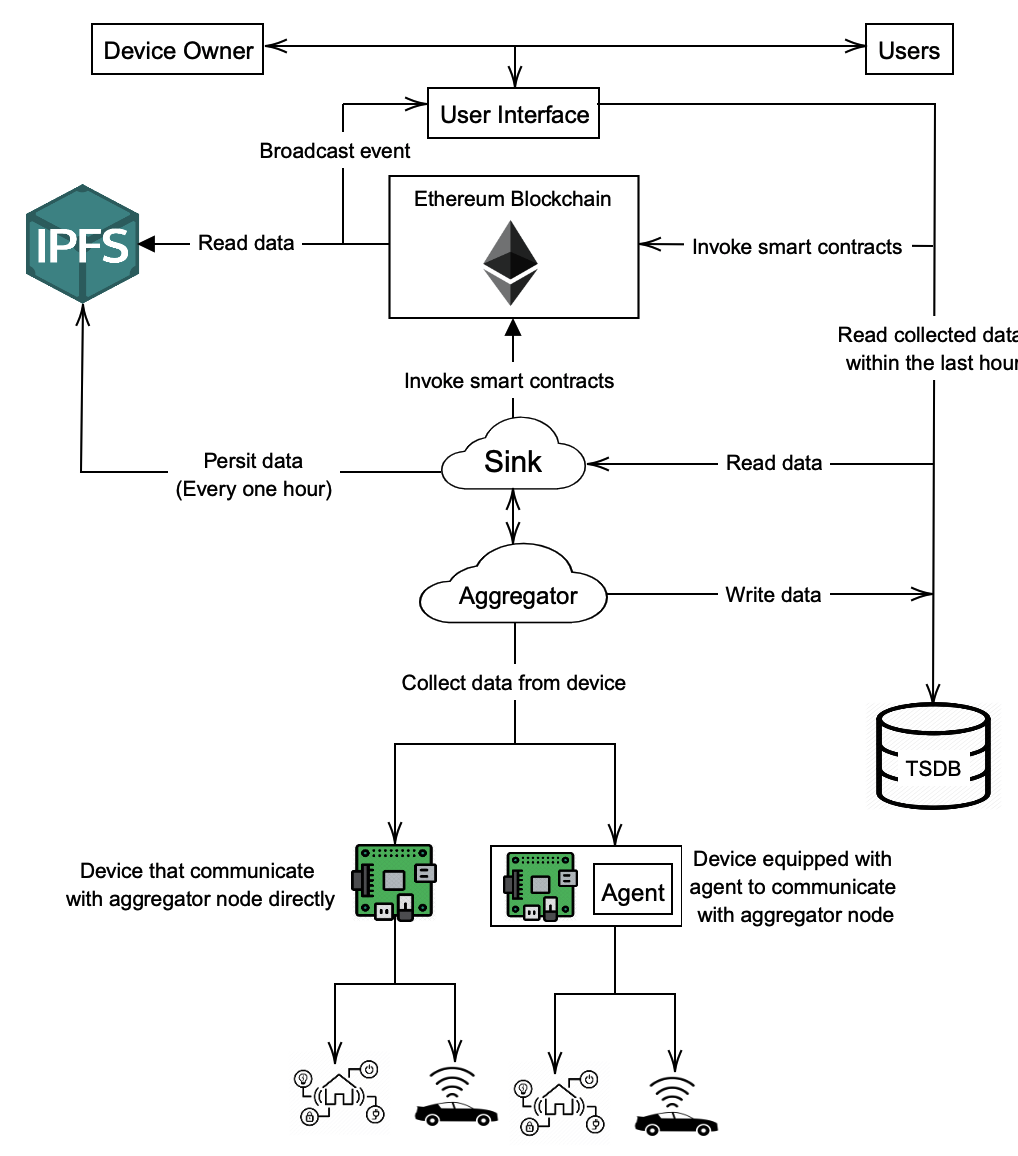}
    \caption{System Architecture}
    \label{fig:conceptualModel}
\end{figure*}

Figure~\ref{fig:conceptualModel} depicts a more detailed model of the system. The main components of the proposed architecture are:
\begin{itemize}
    \item Blockchain network and smart contracts
    \item IPFS file system 
    \item Aggregator nodes
    \item Sink nodes
    \item Gateway devices
    \item Wireless sensor networks
\end{itemize}

These components are described as follows:
\begin{enumerate}
    \item \textbf{Blockchain network and smart contracts}:
    In the proposed architecture, we consider using the Ethereum blockchain network. Smart contracts are created and deployed on the mentioned platform.
    
    \item \textbf{IPFS file system}:
    Due to the fact that storing data directly on the blockchain network is expensive and time-consuming, we have used IPFS~\cite{ipfs} (Interplanetary File System), a decentralized file storage system to store the huge amount of data generated by IoT devices. Several studies (including \cite{ipfs_1},\cite{ipfs_2}, and \cite{ipfs_3}) have leveraged the benefits of IPFS and have built their decentralized applications on top of a software stack of blockchain and IPFS.

    \item \textbf{Fog nodes}:
    Since fog computing brings down the computation to IoT devices, it is considered as a promising technology to help to reduce latency, minimize bandwidth, improve scalability, and enhance agility~\cite{introduction_fogComputing}. Hence, we have employed a layer of fog nodes in our architecture.
    
    In the proposed architecture, fog nodes are divided into two types: sink nodes and aggregator nodes. 

    Sink nodes are considered to have sufficient resources to belong to the blockchain network. Hence, these nodes are capable of taking part in resource-intensive activities and having direct communication with blockchain network.

    Aggregator nodes, in comparison with sink nodes, are assumed to have more constraints in terms of storage and computation resources. These types of nodes are used due to the fact that most of devices that are used in IoT networks are not as resourceful as sink nodes to have direct communication with blockchain.

    \item \textbf{Gateway devices}:
    Gateway devices are connection points between sensors and aggregator nodes. These devices collect data from sensors and preprocess them and finally send them to aggregator nodes. Due to the fact that these devices act as a mediator between aggregator nodes and sensors, an agent can be installed on them to make this communication integrated.
    
    \item \textbf{Wireless sensor networks}:
    Wireless sensor networks (WSNs) are small networks consist of interconnected sensor nodes that communicate wirelessly to collect data about the surrounding environment. These nodes generally have serious limitations in terms of processing and computing power, bandwidth, storage capacity, battery life, and etc. Obviously, due to these limitations, these nodes cannot be part of the blockchain network.
\end{enumerate}

\subsection{System Interactions}
\label{systemInteractions}

This section describes different interactions between different components of the proposed architecture. The interactions can be divided into the following types:

\begin{itemize}
    \item Device registration: Add, edit, view, and delete a device
    \item Threshold definition: Add, edit, view, and delete a monitoring policy
    \item Data collection and event generation: Collect data from IoT devices, store them, and generate appropriate events
    \item Visualization: View collected data and generated events for a device
\end{itemize}

\begin{figure*}[hbt!]
    \begin{subfigure}{.4\textwidth}
        \centering
        \begin{adjustbox}{width=.9\linewidth}
            \includegraphics{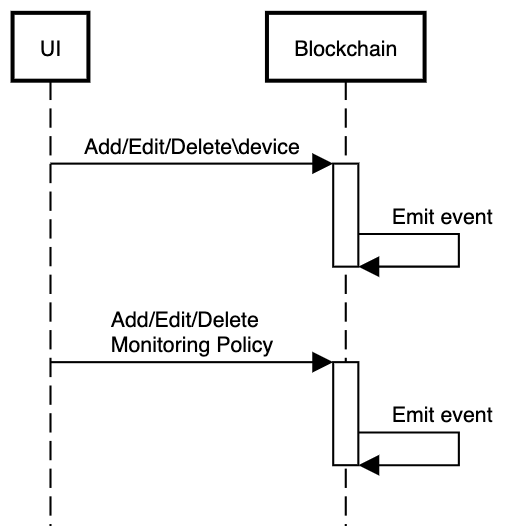}
        \end{adjustbox}
        \caption{Device registration and threshold definition}
        \label{fig:sequenceDiagram1}
    \end{subfigure}
    \begin{subfigure}{.6\textwidth}
        \centering
        \begin{adjustbox}{width=.9\linewidth}
          \includegraphics{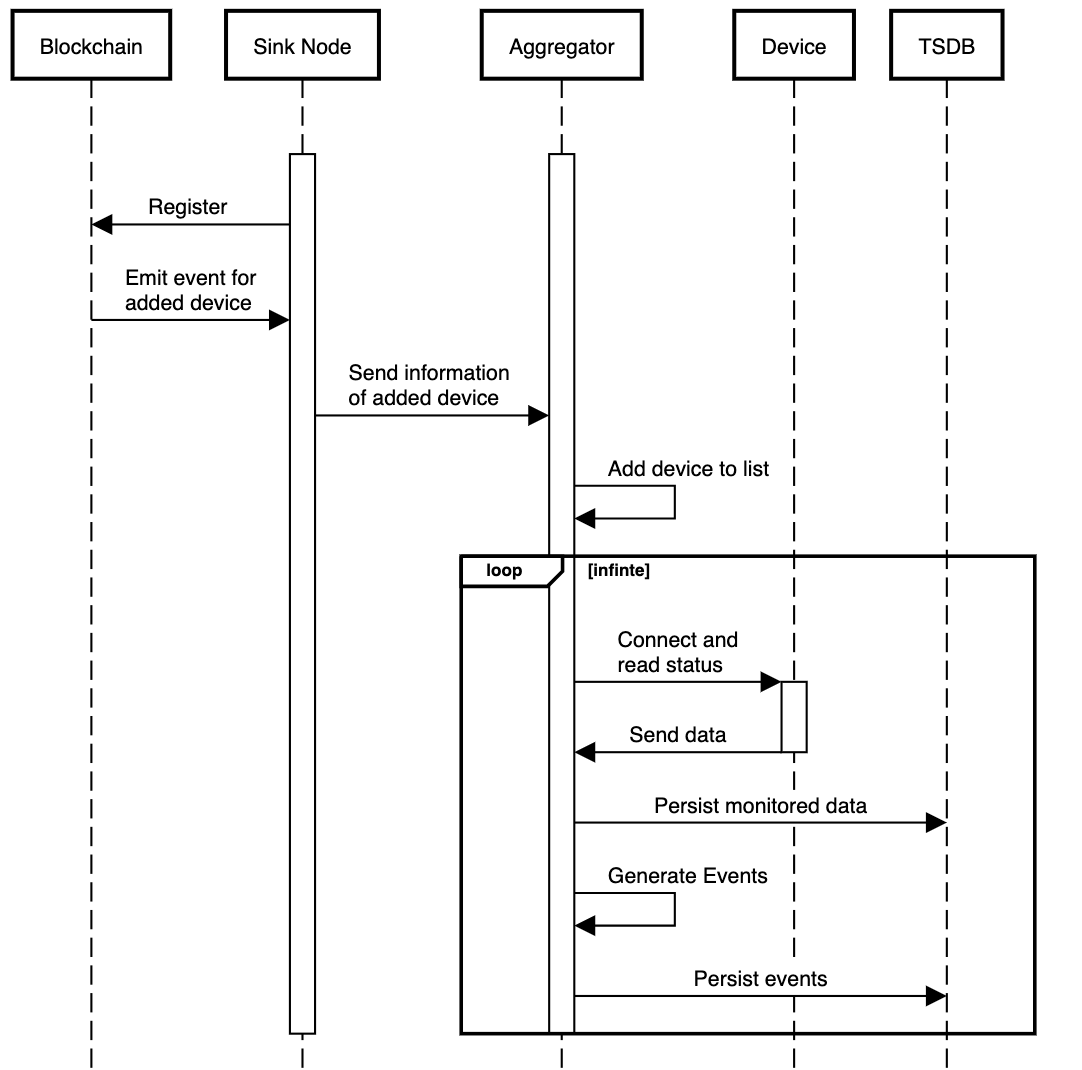}
        \end{adjustbox}
        \caption{Data collection and event generation - Collecting data by aggregator nodes}
        \label{fig:sequenceDiagram2}
    \end{subfigure}

    \begin{subfigure}{.45\textwidth}
        \centering
        \begin{adjustbox}{width=.9\linewidth}
          \includegraphics{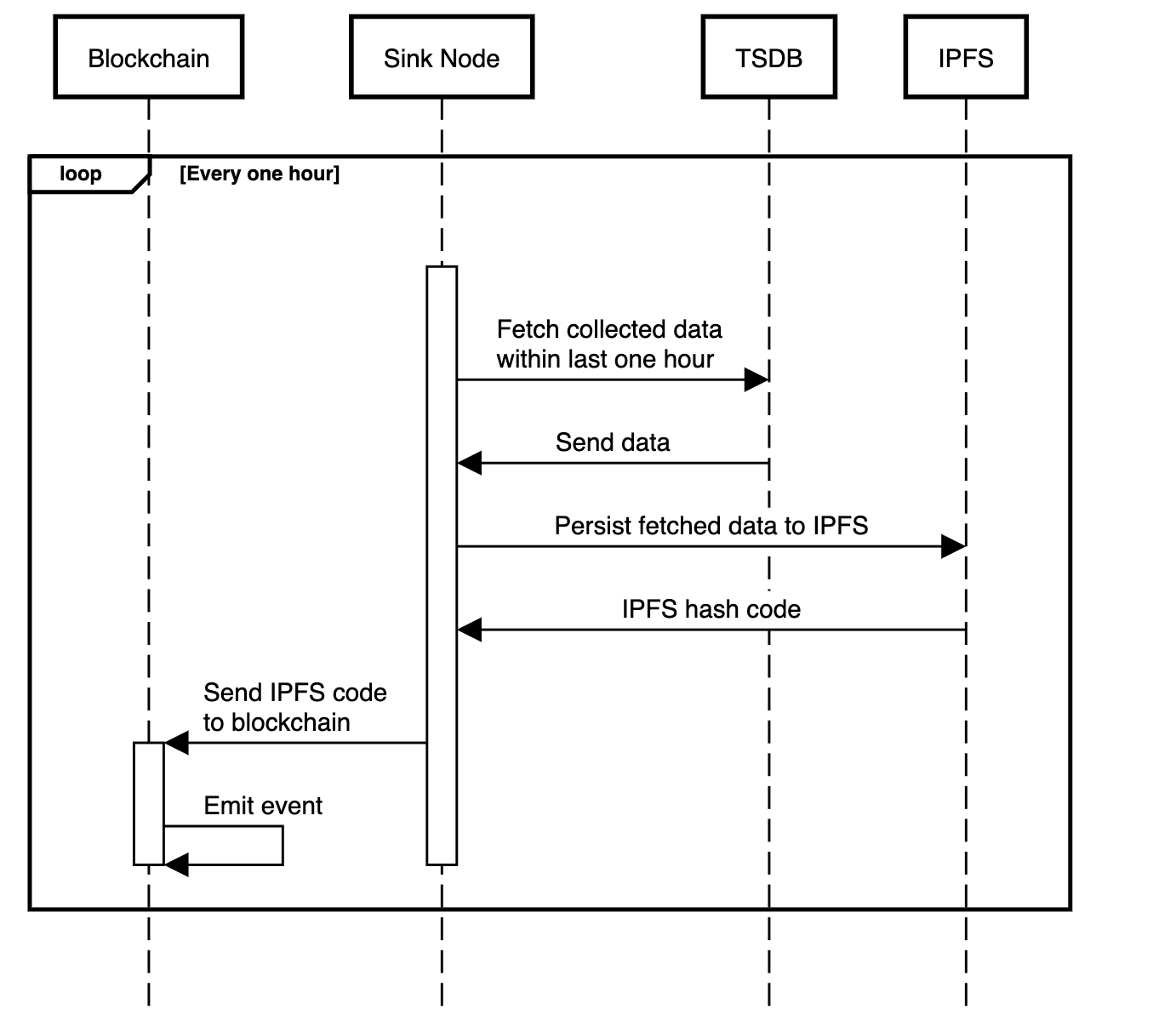}
        \end{adjustbox}
        \caption{Data collection and event generation - Storing collected data by sink nodes}
        \label{fig:sequenceDiagram3}
    \end{subfigure}
    \begin{subfigure}{.55\textwidth}
        \centering
        \begin{adjustbox}{width=\linewidth}
          \includegraphics{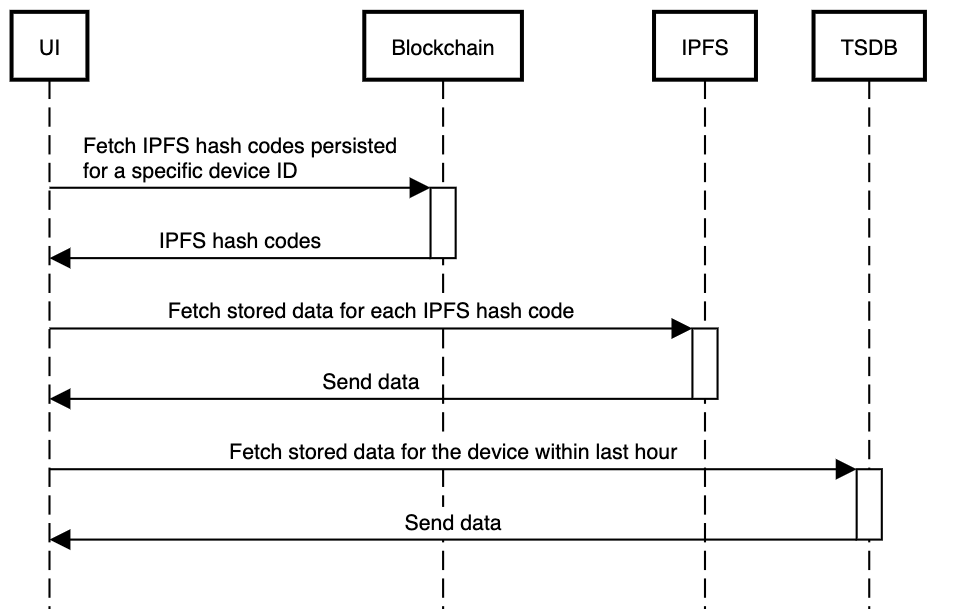}
        \end{adjustbox}
        \caption{Visualization interaction type}
        \label{fig:sequenceDiagram4}
    \end{subfigure}

\caption{Sequence diagram for different interaction types}
\label{fig:sequenceDiagrams}
\end{figure*}

The first two interaction types are depicted in Figure~\ref{fig:sequenceDiagram1}. As is shown in this diagram, two components are involved in this scenario: user interface web console and Ethereum blockchain network. 
To add or modify a device, a transaction should be sent to the blockchain and validated by miners. Hence, adding or modifying a device takes some time to be confirmed. Similarly, adding or modifying a monitoring policy also takes some time to be confirmed.
However, to view a device or policy, a fetch operation should be triggered. As this operation should not be mined, it is executed immediately. 


The third interaction type, data collection and event generation, can be divided into two sub-interactions: collecting data by aggregator nodes and storing them via sink nodes. These two sub-interactions are represented in Figure~\ref{fig:sequenceDiagram2} and Figure~\ref{fig:sequenceDiagram3}, respectively. According to Figure~\ref{fig:sequenceDiagram2}, the process to collect data from IoT devices is as follows:

\begin{enumerate}
    \item 
    As mentioned earlier, sink nodes are resourceful devices that are in direct communication with Ethereum blockchain. They listen to events of adding a new device or monitoring policy. Therefore, when a new device is added to the Ethereum blockchain, the sink node that is listening to that event will be notified.
    
    \item
    Then, the sink node receives the new device information and sends it to a data aggregator node. This information includes but is not limited to: device IP address, its model, its credentials, desired polling interval, and target attributes to be collected.

    \item
    The aggregator node holds a list of devices. In an infinite loop, for each device in that list, the aggregator node polls it and collects its data. Collected data is then persisted in TSDB. Furthermore, they are analyzed to generate events. It is worth mentioning that relevant events are also stored in TSDB.

\end{enumerate}

Then the next step is to persist collected data in Ethereum blockchain. The procedure to transfer TSDB data to blockchain is depicted in Figure~\ref{fig:sequenceDiagram3}. According to Figure~\ref{fig:sequenceDiagram3}, sink nodes fetch data from the TSDB per device and store them into IPFS. The hash code of the stored data will be sent to the Ethereum blockchain. This procedure should be executed at every given time-intervals. It is worth noting that, increasing the length of the time-interval between each execution leads to boosting throughput and storage efficiency. However, it comes at the expense of reducing real-time characteristics of provenance data. 
According to~\cite{ipfs_3}, it is assumed that the length of the time-interval between each execution is 60 minutes. Thus, for each device, the collected data within an hour is persisted on IPFS and its relevant hash code is added to Ethereum blockchain. Since the transaction emitted to add IPFS hash to Ethereum blockchain should be validated by miners, it takes some time to be confirmed.



Finally, the last interaction type is displayed in Figure~\ref{fig:sequenceDiagram4}. As mentioned earlier, for each device, the collected data within each hour, except the last hour, is stored as an IPFS hash code in the Ethereum blockchain. Thus, the steps to read collected monitoring data and generated events for each device are as follows:

\begin{enumerate}
    \item 
    IPFS hash codes for each hour (except the last hour) is fetched from Ethereum blockchain. It worths mentioning that, since this operation should not be mined, it is executed immediately.
        
    \item
    For each fetched IPFS hash code, the content is fetched from IPFS.
    
    \item
    The collected data and events within the last hour is fetched from TSDB.
    
\end{enumerate}


\section{Implementation}
\label{Implementation}

To provide a proof-of-concept, some parts of the proposed architecutre are implemented:

\begin{itemize}
    \item User interface web console
    \item Ethereum blockchain smart contracts
    \item Fog nodes--aggregator and sink nodes
\end{itemize}

Some further details regarding the implementation of each part are described below.

\subsection{User Interface Web Console}
\begin{figure*}[hbt!]
    \begin{subfigure}{.5\textwidth}
        \centering
        \begin{adjustbox}{width=.7\linewidth}
            \includegraphics{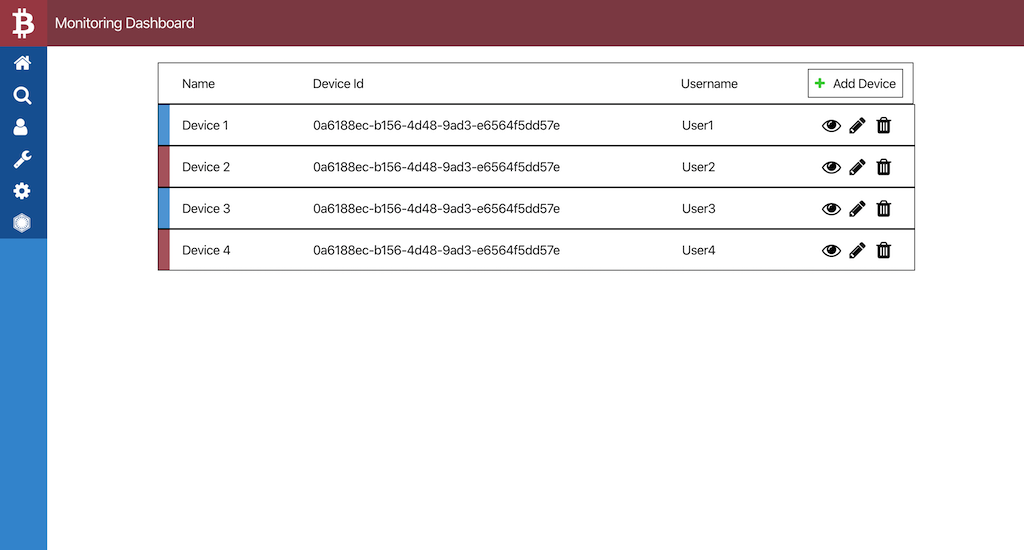}
        \end{adjustbox}       
        \caption{List of devices}
        \label{fig:listOfDevices}
    \end{subfigure}
    \begin{subfigure}{.5\textwidth}
        \centering
        \begin{adjustbox}{width=.7\linewidth}
          \includegraphics{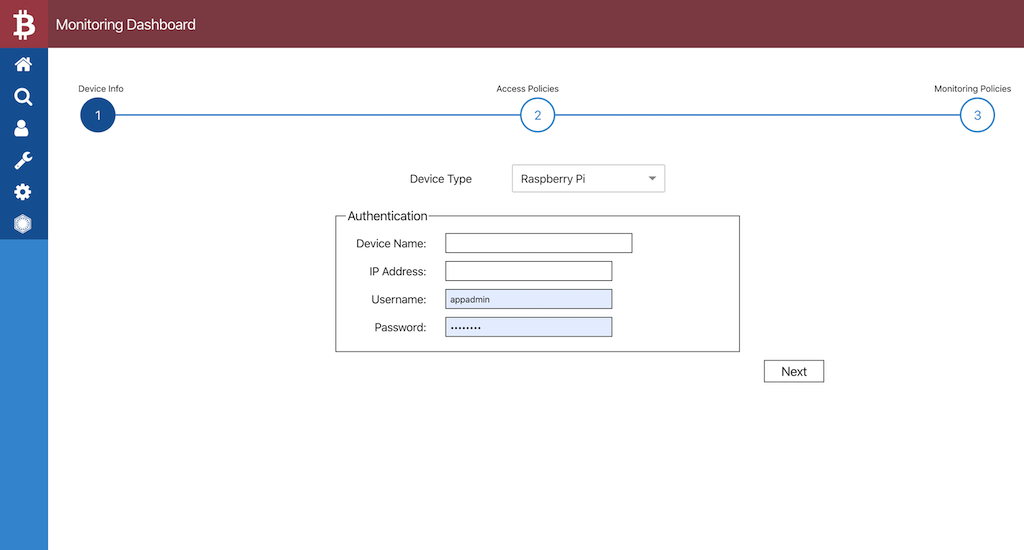}
        \end{adjustbox}
        \caption{Add new device}
        \label{fig:addDevice}
    \end{subfigure}

    \begin{subfigure}{.5\textwidth}
        \centering
        \begin{adjustbox}{width=.7\linewidth}
          \includegraphics{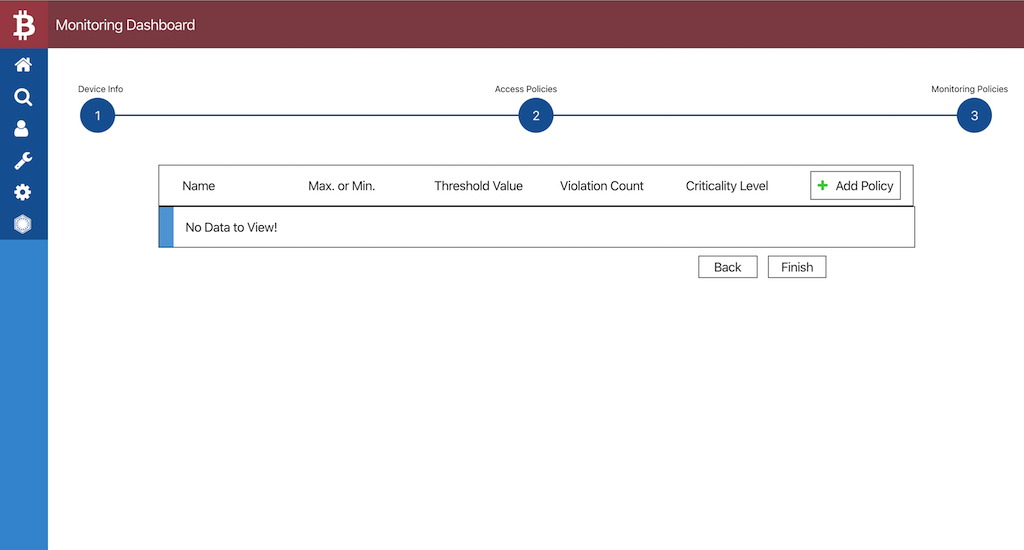}
        \end{adjustbox}
        \caption{List of monitoring policies}
        \label{fig:listOfMonitoringPolicies}
    \end{subfigure}
    \begin{subfigure}{.5\textwidth}
        \centering
        \begin{adjustbox}{width=.7\linewidth}
          \includegraphics{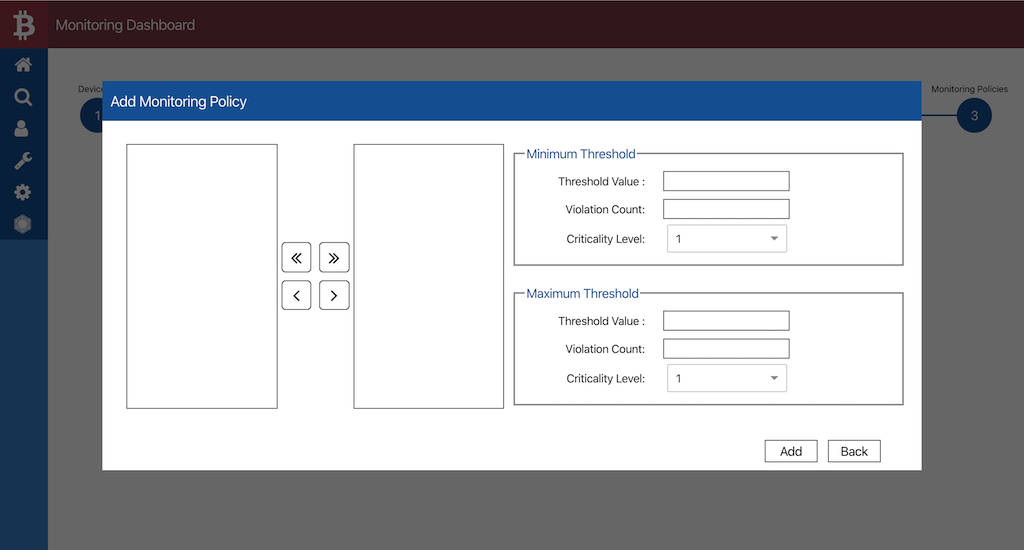}
        \end{adjustbox}
        \caption{Add monitoring policy}
        \label{fig:addMonitoringPolicy}
    \end{subfigure}

\caption{Screenshots of the designed user interface web console}
\label{fig:implementation_webConsole_screenshots}
\end{figure*}

The UI web console application is the entry point for the users to interact with the system and invoke smart contracts. The UI is implemented in javascript, using the ReactJs framework. Furthermore, to be able to interact with UI web console through browsers, without the need to run a full node on machine, Metamask~\cite{metamask} (an extension available on Chrome and Firefox) is used. Moreover,  we adopted the Web3Js, the official Ethereum Javascript API, to communicate with the blockchain. Figure~\ref{fig:implementation_webConsole_screenshots} represents some screenshots of the designed web console.

\subsection{Smart Contracts}
We have implemented our smart contracts in the Solidity programming language~\cite{solidity}, using Remix IDE~\cite{remix}--a browser-based Solidity compiler and IDE. Moreover, we leveraged Geth~\cite{geth} to create our own private Ethereum blockchain and adopted the Truffle framework~\cite{truffle} to deploy our smart contracts. 

Our proposed solution consists of several smart contracts, including \textbf{Monitoring Policies Contract (MPC)} and \textbf{Device Profile Contract (DPC)}, which are respectively used to store monitoring policy attributes and registered devices information. We provide further details on each of these contracts in the following sections.

\subsubsection{Monitoring Policies Contract~(MPC)}
The primary role of MPC is to store monitoring policies for each device. Hence, the MPC maintains a key-value map to store the policies, wherein the key is the device id, and the value is the policy list. Table~\ref{tab:policy_example} illustrates an example of the policy list, in which each row includes the following fields:
\begin{itemize}
    \item Attribute: The attribute for which the policy is defined.
    
    \item Threshold type: Two predefined values are used for threshold type: \textit{Maximum} and \textit{Minimum}. If the policy defines an upper bound for the attribute, the threshold type is set as \textit{Maximum}. Otherwise, in case the policy describes a lower bound, it is set as \textit{Minimum}.

    \item Threshold value: The selected bound for an attribute

    \item Maximum number of violations: This is used as an upper bound for the allowed number of violations. If the number of violations exceeds this limit, a proper event, based on the criticality level, will be emitted.

    \item Criticality level: Different predefined criticality levels, such as \textit{Low}, \textit{Medium}, and \textit{High}
\end{itemize}

\begin{table*}[hbt!]
\centering
\caption{Illustration of policy list}
\label{tab:policy_example}
\begin{tabular}{|c|c|c|c|c|}
\hline
Attribute & Threshold Type & Threshold Value & Max \# of Violations & Criticality Level \\ \hline
Attribute A & $Minimum$ & 10 & 5 & $Medium$ \\ \hline
Attribute B & $Maximum$ & 100 & 10 & $High$ \\ \hline
... & ... & ... & ... & ... \\ \hline
\end{tabular}
\end{table*}

\subsubsection{Device Profile Contract~(DPC)}
We use the DPC to store information and collected statistics, for each device. To achieve this purpose, the DPC maintains a set of properties, called profile, for each device. Table~\ref{tab:deviceProfile} represents the structure of device profiles. As is shown, each profile consists of the following properties:
\begin{itemize}
    \item Device id: A unique identifier for the device
    \item Device model: The model of the device
    \item Polling interval: The polling interval determines how frequently the device should be polled for statistics gathering.
    \item Monitoring data: It is a list of IPFS hash codes for collected monitoring data. As mentioned previously, it is assumed that, for each device, the collected monitoring data is persisted in the blockchain once within an hour. Therefore, each item of the list refers to the collected statistics within one hour.
    \item Events: It is a list of IPFS hash codes for generated events. Similar to the monitoring data list, each item of the list refers to the generated events within one hour.
\end{itemize}

\begin{table*}[hbt!]
\caption{Illustration of device profile}
\label{tab:deviceProfile}
\centering
\begin{tikzpicture}
\node (main) at (0,0){
\begin{tabular}{|p{2.5cm}|}
\hline
Device Id \\ \hline
Device IP Address \\ \hline
Device Model \\ \hline
Polling Interval \\ \hline
\end{tabular}};
\node[yshift=-.065cm] (monitoring) at (main.south){
\begin{tabular}{|p{2.5cm}|}
\hline
Monitoring Data \\ \hline
\end{tabular}};
\node[xshift=3cm] (monitoringHash) at (monitoring.east) 
{
\begin{tabular}{|c|c|c|c|}
\hline
Hash 1 & Hash 2 & Hash 3 & \dots \\ \hline
\end{tabular}
};
\node[yshift=-.065cm] (events) at (monitoring.south){
\begin{tabular}{|p{2.5cm}|}
\hline
Events \\ \hline
\end{tabular}};
\node[xshift=3cm] (eventsHash) at (events.east) 
{
\begin{tabular}{|c|c|c|c|}
\hline
Hash 1 & Hash 2 & Hash 3 & \dots \\ \hline
\end{tabular}
};
\draw[->,ultra thick](monitoring)--(monitoringHash);
\draw[->,ultra thick](events)--(eventsHash);
\end{tikzpicture}
\label{my-label}
\end{table*}

\subsection{Fog Nodes}
As mentioned previously, fog nodes are divided into two types: sink nodes and aggregator nodes. We have implemented both of them in javascript, using NodeJs framework. Moreover, we adopted Web3Js to communicate with the blockchain.

\section{Evaluation}
\label{evaluation}

This section provides further details 
on the experiments we conducted to gain insight into the feasibility of our proposed solution. Section~\ref{evaluation_setups} describes experimental setups more precisely. Performance analysis, scalability analysis, and cost analysis are described in Section~\ref{evaluation_performance}, Section~\ref{evaluation_scalability}, and Section~\ref{evaluation_cost}, respectively. To conduct the evaluation, as is shown in Table \ref{tab:interactions}, we listed all of the possible interactions between users and our proposed framework. For each interaction, a brief description is provided in the second column. Also, the interaction type, in terms of writing to or reading from the blockchain, is mentioned in the third column.
 
It is worth mentioning that writing data to the blockchain emits a transaction that needs to be confirmed by the nodes. Therefore, it is not expected to have a reasonable response time for $write$ interactions due to the high latency of validating transactions. As opposed to transactional interactions, reading data from blockchain does not need to be confirmed by the blockchain nodes. Hence, it is expected that $read$ interactions to be executed in a reasonable time. For this reason, we have limited our performance evaluation to $read$ interactions.

From all interactions listed in Table~\ref{tab:interactions}, the first three rows are $write$ interactions and the last two rows are $read$ interactions. Hence, transactions 1, 2, and 3, in contrast to transactions 4 and 5, are not investigated in Section~\ref{evaluation_cost}. 

\begin{table*}[hbt!]
\centering
\caption{List of all possible interactions}
\label{tab:interactions}
\begin{tabular}{|c|l|c|}
\hline
Number & Description & Interaction Type \\ \hline
1 & Add, edit, and delete a device & Write \\ \hline
2 & Add, edit, and delete a monitoring policy & Write \\ \hline
3 & Add IPFS hash code of monitored data and generated events to the blockchain & Write \\ \hline
4 & Fetch IPFS hash codes of monitored data and generated events from the blockchain & Read \\ \hline
5 & Fetch monitoring policies from the blockchain & Read \\ \hline

\end{tabular}
\end{table*}

\subsection{Experimental Setups}
\label{evaluation_setups}
To implement and execute the proposed solution, a laptop with Mac OS, Core i5 2.9 GHz, and 8GB memory is used.

Furthermore, to be able to simulate the real-world transactions, we have used the Ropsten network, a public Ethereum test network, as our testbed. Hence, all metrics which are described in the following sections are collected on Ropsten testbed.

We have generated 200 devices and created four benchmarks: In the first benchmark, which is called B1, 50 devices are concurrently active. In the second one, named as B2, 100 devices are simultaneously active. In the third one, B3, 150 devices are concurrently active. Finally, in the last benchmark, which is referred to as B4, all 200 devices are simultaneously active. In all benchmarks, the rate of data generation and polling interval for all devices are the same.

\subsection{Performance Analysis}
\label{evaluation_performance}

As mentioned before, there are two types of interactions between a device and our framework: the interactions that write data by triggering blockchain transactions, and those that query to read data. The former needs to be validated by the miners and, consequently, incurs high delays and is not expected to be executed in a reasonable time. However, the latter does not incur any significant delay and is expected to be executed in a reasonable time. Hence, in this section, we investigate $read$ interactions (interactions number 4 and 5 from Table~\ref{tab:interactions}) by measuring the performance metrics.  

To evaluate the performance of $read$ interactions, response time is measured and reported in Figure~\ref{chart:responseTime}. The approach of using response time as a metric for evaluating performance is used in several papers, including \cite{evaluation_performance_responseTime_1} and \cite{ipfs_1}.

\definecolor{bblue}{HTML}{4F81BD}
\definecolor{rred}{HTML}{C0504D}
\definecolor{ggreen}{HTML}{9BBB59}
\definecolor{ppurple}{HTML}{671d9d}

\begin{figure*}[!hbt]
    \centering
    \begin{tikzpicture}
        \centering
        \begin{axis}[
            ybar, axis on top,
            title={},
            height=4cm, width=\textwidth,
            bar width=0.9cm,
            ymajorgrids, tick align=inside,
            major grid style={draw=white},
            enlarge y limits={value=.1,upper},
            ymin=0, ymax=2500,
            axis x line*=bottom,
            axis y line*=left,
            y axis line style={opacity=0},
            tickwidth=0pt,
            enlarge x limits=true,
            area legend,
            legend style={
                at={(0.5,1.5)},
                anchor=north,
                legend columns=-1,
                /tikz/every even column/.append style={column sep=0.2cm}
            },
            ylabel={Average response time (ms)},
            xlabel={Benchmarks},
            symbolic x coords={B1,B2,B3,B4},
            xtick=data,
            yticklabels=\empty,
            nodes near coords={
                \pgfmathprintnumber[precision=1]{\pgfplotspointmeta}
            }
        ]
        
        \addplot [draw=none, fill=bblue, postaction={pattern=vertical lines}] 
        coordinates {(B1, 1523)(B2, 1702)(B3, 1757)(B4, 1699)};
        
        \addplot [draw=none, fill=rred, postaction={pattern=horizontal lines}] coordinates {(B1, 1413)(B2, 1547)(B3, 2197)(B4, 1943)};
        
        \legend{Fetch monitoring policies, Fetch IPFS hash codes}
    
        \end{axis}
    \end{tikzpicture}
    \caption{Average response time for all benchmarks}
    \label{chart:responseTime}
\end{figure*}
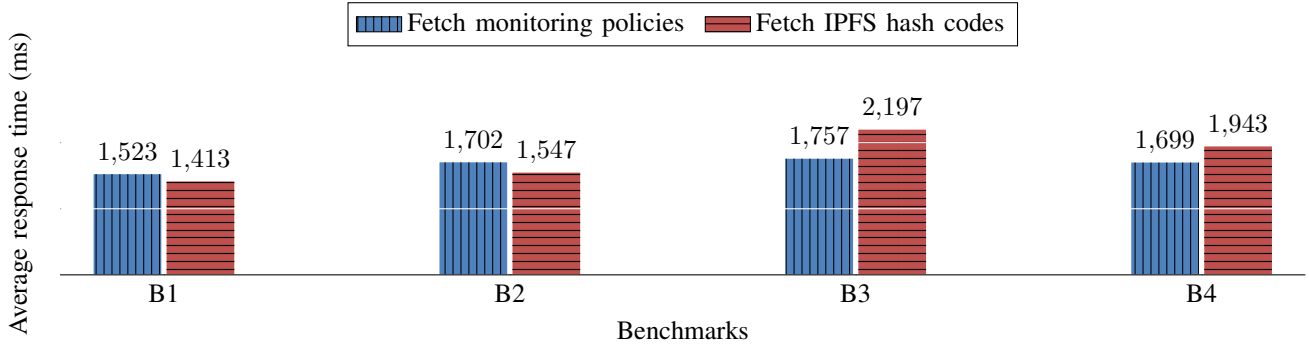

\subsection{Scalability Analysis}
\label{evaluation_scalability}
To evaluate the scalability of our solution, two aspects are considered to be of paramount importance. The first is that, due to the resource-constrained nature of IoT devices, it is not feasible to store the whole blockchain data on them. The second is the technical limitation regarding the throughput of transactions. This limitation is caused due to the difficulty of the Proof-of-Work algorithm, which is used as the consensus algorithm in Ethereum 1.x \cite{ethereum_whitePaper}.

To address the first problem, we have used fog nodes, which are assumed to be resource-intensive devices and are capable of storing the whole blockchain. Therefore, the IoT device does not need to store the blockchain anymore.

Apart from that, our approach to address the second problem is to minimize the number of transactions sent to the Ethereum blockchain. To do so, we considered that the monitored data is persisted in TSDB. Moreover, at regular and specific intervals, the collected data for each device are batched and stored to IPFS, and then, the generated IPFS hash codes for all devices are packed and sent to the Ethereum blockchain. Another question then arises: As each IPFS hash code represents a device, how many devices can be handled in each transaction? To answer this question, 
let us denote the gas limit for each transaction by $GasLimit$, the gas paid for each transaction by $G_{transaction}$, the gas paid for every non-zero byte of data for each transaction by $G_{txdatanonzero}$, and the IPFS hash code size by $HashSize$. As of writing this paper, according to \cite{ethereum_yellowPaper} and \cite{ethereum_gasLimit}, in the Ethereum network, $GasLimit$, $G_{transaction}$, and $G_{txdatanonzero}$ are equal to \numprint{6500000}, \numprint{21000}, and 68, respectively. Additionally, as mentioned before, in our proposed solution, it is assumed that collected data is persisted in the blockchain every one hour. Hence, the maximum number of devices that can be handled per transaction in our proposed solution can be calculated as follows:
\begin{equation}
\begin{aligned}
\label{equation:handledDevices}
\left \lfloor \frac{GasLimit - G_{transaction}}{G_{txdatanonzero} \times HashSize} \right \rfloor = 2977
\end{aligned}
\end{equation}

Therefore, according to Equation~\ref{equation:handledDevices}, the maximum number of devices that can be handled per transaction is approximately 2977. In another term, for every 2977 devices, only one transaction is sent to the Ethereum blockchain every hour, which is a piece of evidence for the scalability of our framework.

\subsection{Cost Analysis}
\label{evaluation_cost}

In the Ethereum network, sending any write operation, whether it succeeds or fails, needs to be validated and executed by miners, which takes computational power. Hence, any write operation incurs some amount of fee, which is determined regarding $Gas$ and $Gas Price$. \cite{ethereum_whitePaper}

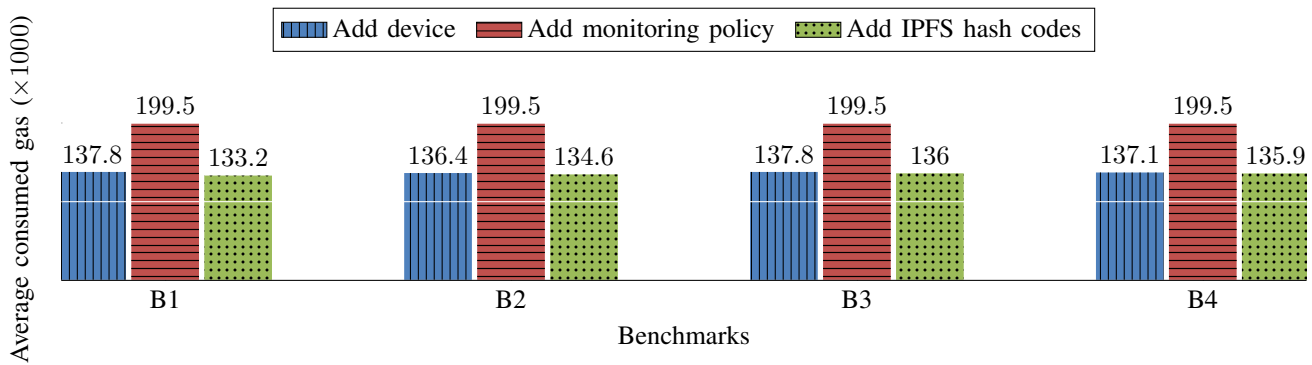
\begin{figure*}[!hbt]
    \centering
    \begin{tikzpicture}
        \centering
        \begin{axis}[
            ybar, axis on top,
            title={},
            height=4cm, width=\textwidth,
            bar width=0.9cm,
            ymajorgrids, tick align=inside,
            major grid style={draw=white},
            enlarge y limits={value=.1,upper},
            ymin=0, ymax=210,
            axis x line*=bottom,
            axis y line*=left,
            y axis line style={opacity=0},
            tickwidth=0pt,
            enlarge x limits=true,
            area legend,
            legend style={
                at={(0.5,1.5)},
                anchor=north,
                legend columns=-1,
                /tikz/every even column/.append style={column sep=0.2cm}
            },
            ylabel={Average consumed gas  ($\times 1000$)},
            xlabel={Benchmarks},
            symbolic x coords={B1, B2, B3, B4},
            xtick=data,
            yticklabels=\empty,
            nodes near coords={\pgfmathprintnumber[precision=1]{\pgfplotspointmeta}}
        ]
        
        \addplot [
            draw=none, 
            fill=bblue,
            postaction={
                pattern=vertical lines
            }
        ] coordinates {(B1, 137.780)(B2, 136.402)(B3, 137.779)(B4, 137.091)};
  
        \addplot [
            draw=none,
            fill=rred,
            postaction={
                pattern=horizontal lines
            }
        ] coordinates {(B1, 199.464)(B2, 199.464) (B3, 199.464)(B4, 199.465)};
        
        \addplot [
            draw=none,
            fill=ggreen,
            postaction={
                pattern=dots
            }
        ] coordinates {(B1, 133.176)(B2, 134.642)(B3, 136.047)(B4, 135.917)};
        
        \legend{Add device,Add monitoring policy,Add IPFS hash codes}
    
        \end{axis}
    \end{tikzpicture}
    \caption{Average gas used for all benchmarks}
    \label{chart:gas}
\end{figure*}

Figure~\ref{chart:gas} shows the average gas spent over each type of transactions when run on the Ropsten network. To gain an understanding of how cost-effective our solution is, according to \cite{evaluation_cost_gasConversionToDollar_1} and \cite{evaluation_cost_gasConversionToDollar_2}, we have converted the amount of gas used per transaction type (average over four benchmarks) to the US dollar. To do so, we multiplied the consumed gas with the default Truffle gas price (1 gas =  100 Gwei)~\cite{truffleGasPrice}. The results are reported in Table \ref{tab:transactionsGasUsed}.

\begin{table*}[hbt!]
\centering
\caption{Transaction cost by US dollar (Average over four benchmarks)}
\label{tab:transactionsGasUsed}
\begin{tabular}{|c|l|c|c|c|}
\hline
Number & Description & Average Gas & $Gas \times Gas Price$ & Average Cost (\$) \\ \hline
1 & Add, edit, and delete a device & 137.2 & 13720 Gwei & $\approx 0.0018$ \$ \\ \hline
2 & Add, edit, and delete a monitoring policy & 199.5 & 19950 Gwei & $\approx 0.0026$ \$ \\ \hline
3 & Add IPFS hash code of monitored data and generated events to the blockchain & 134.6 & 13460 Gwei & $\approx 0.0018$ \$ \\ \hline
\end{tabular}
\end{table*}

According to Table \ref{tab:transactionsGasUsed} and Figure~\ref{chart:gas}, it can be concluded that among all transaction types, adding a monitoring policy is the most expensive one. This can be an evidence that the transaction for adding a monitoring policy consumes more computational power. However, the results demonstrate that the solution is cost-efficient.



\subsection{Limitations}
Despite the contributions put forth by this paper, the proposed solution suffers from some limitations. One limitation is that it is assumed the procedure of persisting data to the blockchain is repeated every one hour. However, increasing the period duration with the goal of boosting throughput and storage efficiency, comes at the expense of reducing real-time characteristics of provenance data. In other terms, there is a trade-off between throughput and being real-time, which needs more investigation. Another limitation is that due to the immaturity of the blockchain technology, there are few studies of using blockchain to monitor IoT devices. Moreover, most of these studies are theory-based and lack providing implementation details.
Furthermore, as to the best of our knowledge, it is the first study that considers adding dynamic policies for monitoring IoT devices, no similar study was found to be referred to as a benchmark.

\section{Conclusion}
\label{Conclusion}
The integration of blockchain technology into IoT systems is in its infancy. However, in recent years, several researchers explored the benefits, challenges, and use-cases of the combination of blockchain and IoT. In this paper, we proposed a novel scalable architecture based on Ethereum blockchain for IoT device monitoring. Moreover, to enhance the scalability of our solution, we have used fog nodes and IPFS file system.

\bibliographystyle{unsrt}
\bibliography{references}

\end{document}